\documentclass[twocolumn,aps,prc,superscriptaddress,showpacs]{revtex4}
\usepackage{amsmath,bm}%
\usepackage{graphicx}%

\begin{document}

\title{Dynamical dipole $\gamma$ radiation in heavy-ion collisions on the basis of the quantum molecular dynamics model}

\author{ H. L. Wu}
\affiliation{Shanghai Institute of Applied Physics, Chinese
Academy of Sciences, P.O. Box 800-204, Shanghai 201800, China}
\affiliation{Graduate School of the Chinese Academy of Sciences,
Beijing 100080, China}

\author{ W. D. Tian}
\affiliation{Shanghai Institute of Applied Physics, Chinese
Academy of Sciences, P.O. Box 800-204, Shanghai 201800, China}

\author{ Y. G. Ma}
\thanks{Corresponding author. Email: ygma@sinap.ac.cn}
\affiliation{Shanghai Institute of Applied Physics, Chinese
Academy of Sciences, P.O. Box 800-204, Shanghai 201800, China}

\author{ X. Z. Cai}
\affiliation{Shanghai Institute of Applied Physics, Chinese
Academy of Sciences, P.O. Box 800-204, Shanghai 201800, China}

\author{ J. G. Chen}
\affiliation{Shanghai Institute of Applied Physics, Chinese
Academy of Sciences, P.O. Box 800-204, Shanghai 201800, China}

\author{ D. Q. Fang}
\affiliation{Shanghai Institute of Applied Physics, Chinese
Academy of Sciences, P.O. Box 800-204, Shanghai 201800, China}

\author{ W. Guo}
\affiliation{Shanghai Institute of Applied Physics, Chinese
Academy of Sciences, P.O. Box 800-204, Shanghai 201800, China}

\author{ H. W. Wang}
\affiliation{Shanghai Institute of Applied Physics, Chinese
Academy of Sciences, P.O. Box 800-204, Shanghai 201800, China}

\date{\today}

\begin{abstract}

Dynamical dipole $\gamma$-ray  emission in heavy-ion collisions is
explored in the framework of the quantum molecular dynamics model.
The studies are focused on systems of $^{40}$Ca bombarding
$^{48}$Ca and its isotopes at different incident energies and
impact parameters. Yields of $\gamma$ rays are calculated and the
centroid energy and dynamical dipole emission width of the
$\gamma$ spectra are extracted to investigate the properties of
$\gamma$ emission. In addition, sensitivities of dynamical dipole
$\gamma$-ray emission to the isospin and the symmetry energy
coefficient of the equation of state are studied. The results show
that detailed study of dynamical dipole $\gamma$ radiation can
provide information on the equation of state and the symmetry
energy around the normal nuclear density.
\end{abstract}

\pacs{25.70.Gh, 24.30.Cz, 25.70.Ef, 25.70.Lm}

\maketitle

Giant dipole resonance (GDR) $\gamma$-ray emission built on
heavy-ion collisions, which  was predicted  in early macroscopic
models \cite{Ho79,Di85} has been widely studied both
experimentally and theoretically during the past few decades
\cite{review,Pa03}. It can  be ascribed to  a collective vibration
of protons against neutrons with a dipole spatial pattern that
starts with the overlap of two  nuclei in both fusion or
incomplete fusion. Recently, a so-called dynamical dipole mode
(pre-equilibrium GDR) that is generated by a high-amplitude dipole
collective motion of nucleons before the formation of a fully
equilibrated compound nucleus (CN) has been discussed
\cite{Si01,Di08,Co09} and it can encode information about the
early stage of collisions when the CN is still in a highly
deformed configuration. Some efforts have been made to study the
dipole resonance formed during fusion in $N/Z$ asymmetry heavy-ion
reactions as it could be a good probe for the symmetry term in the
equation of state (EOS), which is one of the hotly debated issues
related to nuclear astrophysics problems such as the elements
burning in supernova \cite{BALi}.

Several microscopic transport simulations, such as time-dependent
Hartree-Fock  \cite{Si01}, Boltzmann-Nordheim-Vlasov \cite{Ba09},
and constrained molecular dynamics \cite{Pa05}, have been
successfully used to study GDR dynamical dipole emission. However,
the quantum molecular dynamics (QMD) model, which has been
effectively used to dispose problems of heavy-ion collisions up to 2
GeV/nucleon \cite{Ai91}, has not been used to investigate GDR. In
principle, energy is not conserved in the QMD model and the model is
more suited for higher-energy phenomena. But, it is still of
interest to see how the QMD model works for GDR $\gamma$ radiation
and discuss its properties .

We use the isospin-dependent QMD (IDQMD) model, which is based on
the general QMD model but embodies the isospin degrees of freedom in
two-body nucleon-nucleon collisions and Pauli blocking, as well as
considering the difference between neutron and proton density
distributions for nuclei far from the $\beta$-stability line and
treating the sampling of neutrons and protons in the initialization
in phase space, respectively \cite{Wei-Ma,Tian}.

The mean field involved in the present IDQMD model is given by
$U(\rho) = U^{\rm Sky} + U^{\rm Coul}  + U^{\rm Yuk} + U^{\rm
sym}$,
with $U^{\rm Sky}$, $U^{\rm Coul}$, $U^{\rm Yuk}$ ,and  $U^{\rm
sym}$  representing the Skyrme potential, the Coulomb potential, the
Yukawa potential and the symmetry potential interaction,
respectively \cite{Ai91}. The Skyrme potential is 
$U^{\rm Sky} = \alpha(\rho/\rho_{0}) +
\beta{(\rho/\rho_{0})}^{\gamma}$,
where $\rho_{0}$ = 0.16 {fm}$^{-3}$(the saturation nuclear
density) and $\rho$ is the nuclear density. 
 We take the parameters $\alpha =-356$ MeV, $\beta =303$ MeV, and $\gamma =7/6$, corresponding to a
soft EOS, and $\alpha=-124$ MeV, $\beta=70.5$ MeV, and $\gamma = 2$,
corresponding to a stiff EOS. $U^{\rm Yuk}$ is a long-range
interaction that is necessary to describe the surface of the nucleus
and takes the following form:
$ U^{Yuk}  =({V_y}/{2}) \sum_{i \neq j}{exp(Lm^2)}/{r_{ij}}
\cdot[exp(mr_{ij})erfc(\sqrt{L}m-{r_{ij}}/{\sqrt{4L}})-exp(mr_{ij})
erfc(\sqrt{L}m+{r_{ij}}/{\sqrt{4L}})]$
with $V_y = 0.0024$GeV, $m= 0.83$, and $L$ is the so-called Gaussian
wave-packet width (here $L = 2.0$ fm$^{2}$). $r_{ij}$ is the
relative distance. In principle, if one adds $U^{\rm Yuk}$ into the
effective potential, some part of the two-body term in $U^{\rm loc}$
should be subtracted. Therefore $\alpha$ should be smaller
\cite{Ai91}. However, considering that we aim to discuss some
qualitative effects of the stiffness of potential on pre-equilibrium
GDR $\gamma$ emission, we tentatively keep the preceding two sets of
potential parameters.

In recent years, some efforts have been made to constrain the
symmetry energy term of the EOS from nuclear reactions induced by
asymmetric heavy ions, and various probes have been proposed
\cite{BALi}. Considering that the pre-equilibrium GDR $\gamma$ ray
originates from the isospin nonequilibrium of the entrance channel
or the composite intermediate system, its spectra should be
dependent on the symmetry potential. The symmetry potential in the
IDQMD model is obtained by
\begin{equation}
U^{\rm sym} = \frac{C_{\rm sym}}{2\rho_{0}}\sum_{i{\neq}
j}{\tau_{iz}\tau_{jz}\frac{1}{(4{\pi}L)^{3/2}}\exp{[-\frac{(r_{i}-r_{j})^{2}}{4L}]}}
\end{equation}
with $C_{\rm sym}$ the symmetry energy strength, $\tau_{\rm z}$
being the $z$th component of the isospin degree of freedom, which
equals 1 or $-1$ for neutrons or protons, respectively. A detailed
description of the QMD model is given in Ref.\cite{Ai91}.

The giant dipole moment in coordinator space [$DR(t)$] and in
momentum space [$DK(t)$] is written, respectively, as follows
\cite{Ba01}:
\begin{equation}
DR(t) = \frac{NZ}{A}X(t)=\frac{NZ}{A}{(R_{Z}-R_{N})},
\end{equation}
where $X(t)$ is the distance between the centers of mass of protons
and neutrons, $R_{Z}=\sum\limits_{i}{x_{i}(p)}$ and
$R_{N}=\sum\limits_{i}{x_{i}(n)}$ is the center of mass of protons
and neutrons, respectively, and
\begin{eqnarray} &&
DK(t) =
\frac{NZ}{A\hbar}\prod(t)=\frac{NZ}{A\hbar}(\frac{P_{p}}{Z}-\frac{P_{n}}{N})
\end{eqnarray}
is just the canonically conjugate momentum of the $DR(t)$, where
$\prod(t)$ denotes the relative momentum, with $P_{n}$ $(P_{p})$
centers of mass in momentum space for neutrons (protons).

Here  $N$ = $N_{P} + N_{T}$ and $Z = Z_{P} + Z_{T}$, the suffix $P$
(or $T$) marks projectile (or target). We have adopted the initial
dipole moment as follows:
\begin{eqnarray} &&
DR(t=0) = \frac{NZ}{A}|R_{Z}(t=0)-R_{N}(t=0)|\nonumber\\
    &
    &=\frac{R_{P}+R_{T}}{A}Z_{P}Z_{T}|(\frac{N}{Z})_{T}-(\frac{N}{Z})_{P}|,
\end{eqnarray}
where $R_{P}$ and $R_{T}$ are the radii of the projectile and
target, respectively.

Derived from the overall dipole moment $D(t)$, we can get the
$\gamma$-ray emission probability for energy $E$ \cite{Pa03}
\begin{equation}
\frac{dP}{dE}=\frac{2}{3\pi}\frac{e^{2}}{E{\hbar}c}
|\frac{\overline{dV_{k}}}{dt}(E)|^{2},
\end{equation}
where $dP$/$dE$ can be interpreted as the average number of $\gamma$
rays emitted per energy unit, while
$\frac{\overline{dV_{k}}}{dt}(E)$ is the Fourier transformation of
the time derivative on the $k$th components $x$ and $z$ of
$\overline{\overrightarrow{V}}$,
\begin{equation}
\frac{\overline{dV_{k}}}{dt}(E)=\int_{0}^{\infty}{\frac{d\overline{V_{k}}}
{dt}(t)e^{i({Et}/{\hbar})}dt},
\end{equation}
where $\overline{V}$ denotes the dipole resonance of the system and
was derived from the total dipole according to the equation
\begin{equation}
\overline{V_{k}}=\frac{d\overline{D(t)_{k}}}{dt}.
\end{equation}

\begin{figure}
\includegraphics[scale=0.18]{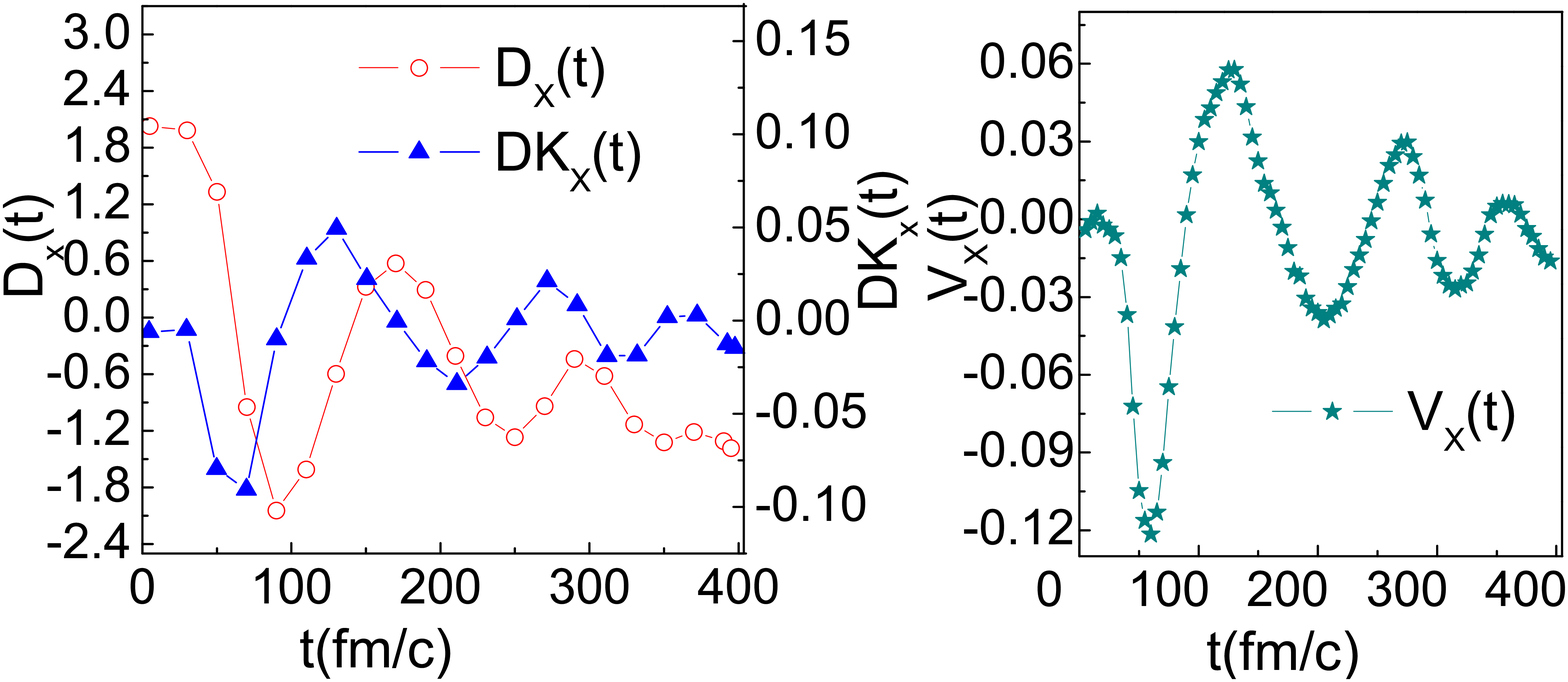}
\vspace{-0.1truein} \caption{\footnotesize (Color online) Left:
Evolution of the giant dipole moment with time in coordinate space
[$D_{x}(t)$] and momentum space [$DK_{x}(t)$]. Right: Dynamical
dipole mode [$V_{x}(t)$]. The reaction system is $^{40}$Ca+$^{48}$Ca
collision at 10 MeV/nucleon with $b$= 1 fm and soft EOS
parameters.}\label{Fig1}
\end{figure}

Figure \ref{Fig1} shows the evolution of the giant dipole moment
with time in coordinate space and momentum space for $^{40}$Ca +
$^{48}$Ca collisions at 10 MeV/nucleon and $b = 1$ fm. The giant
dipole moment attenuates rapidly after $t = 400$ fm/$c$,
illustrating a clear reduction of the collective behavior. The
similar behavior is observed for the pre-equilibrium or dynamical
dipole mode [denoted $V_{x}(t)$].

\begin{figure}
\includegraphics[scale=0.2]{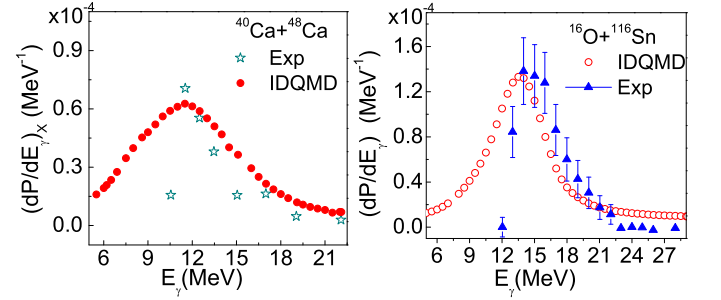}
\vspace{-0.1truein} \caption{\footnotesize (Color online) Left:
Experimental data (stars) \cite{Pa05} together with our IDQMD model
[filled (red) circles] of the $\gamma$-ray spectrum of the
$^{40}$Ca+$^{48}$Ca system at 10 MeV/nucleon. Right: Measured value
of $\gamma$-ray yield for the $^{16}$O+$^{116}$Sn system at 15.6
MeV/nucleon (open circles) \cite{Co09} and our calculation results
[filled (blue) triangles] under the same conditions.}\label{Fig2}
\end{figure}

To determine the reliability of our calculations of dipole emission
with the IDQMD model, experimental data for the $^{40}$Ca+$^{48}$Ca
system \cite{Pa05} at $E_{\rm beam}= 10$ MeV/nucleon and the
$^{16}$O+$^{116}$Sn system \cite{Co09} at $E_{\rm beam} = 15.6$
MeV/nucleon are compared with our results in Fig. \ref{Fig2}. It
should be noted that the experimental data on dynamical dipole
emission here are obtained by taking away the statistical
contribution (results of the statistical model) from the measured
spectra. In addition, calculations for both systems are carried out
with soft Skyrme potential and impact parameters being a triangle
distribution between 1 and 6 fm. The good agreement between the data
and the IDQMD calculation encourages us to make systematic
calculations on pre-equilibrium GDR $\gamma$ emission, such as its
dependences on impact parameter, isospin, EOS, etc.

\begin{figure}
\vspace{-0.1truein}
\includegraphics[scale=0.19]{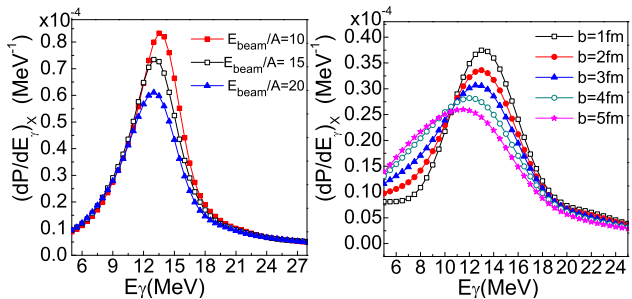}
\vspace{-0.1truein} \caption{\footnotesize (Color online) Left:
Dipole dynamical emission spectra of the $^{40}$Ca$+$$^{48}$Ca
system at 10, 15, and 20 MeV/nucleon, with impact parameters $b=1$
fm and soft EOS parameters. Right: Dynamical dipole $\gamma$-ray
yield distribution of the $^{40}$Ca+$^{48}$Ca system at 10
MeV/nucleon in the early 400 fm/$c$, with impact parameter $b$
varied from 1 to 5 fm and soft EOS parameters. }\label{Fig3}
\end{figure}

So far both experimental data and theoretical calculations have
shown that both the centroid and the width of the pre-equilibrium
dipole component, with the exception of the $\gamma$-ray yield of
GDR, are almost independent of the beam energy, and in this study, a
similar conclusion  was drawn from the left panel in Fig.
\ref{Fig3}, where we report calculations of the dynamical
$\gamma$-ray emission of $^{40}$Ca+$^{48}$Ca at $E_{\rm beam}=$ 10,
15, and 20 MeV/nucleon, respectively. In the figure, the centroid
and width of $\gamma$ spectra seem insensitive to the incident
energy, but the $\gamma$-ray emission probability $dP/dE$ is
suppressed at higher incident energies. A direct reduction of the
initial charge asymmetry owing to pre-equilibrium nucleon emission
(mostly neutrons) may account for the depressed $\gamma$ yield. In
addition, a higher excitation energy deposited in the composite
system can damp the GDR emission. The suppressed behavior of
$\gamma$-ray emission with increasing incident energy means that the
heavy-ion collisions at a relative low energy are suitable for
studying the collective dipole resonance. In fact, the conclusion
here was supported by the existence of the dynamical dipole mode
studied in fusion \cite{Pi,Am04,Fl48} and deep-inelastic heavy-ion
collisions \cite{Pa05,Am04}.

The right panel in Fig. \ref{Fig3} gives the $\gamma$-ray emission
probability of the $^{40}$Ca+$^{48}$Ca system at 10 MeV/nucleon on a
scale of 0$-$400 fm/$c$, with $b$ = 1$-$5 fm and soft EOS
parameters. Both the yield and the central energy of the
$\gamma$-ray distribution show a decreasing tendency with a larger
impact parameter. As we know, a larger impact parameter inevitably
induces a larger time scale of the isospin variation, corresponding
to a low frequency of the dipole resonance, and thus leads to a
lower central energy. Besides, collisions with different impact
parameters, corresponding to different angular momenta, may generate
dinuclei of different geometrical shapes \cite{Bracco}, and
necessarily, a larger impact parameter produces a dinuclei system
with a larger deformation. Using the conclusion in Ref.\cite{Pi},
the GDR in a deformable CN is expected to have a central energy
lower than that in a spherical nucleus of similar mass. Therefore
the suppression of the central energy with an increasing impact
parameter can also be partly attributed to the larger deformation of
the dinuclei at the emission moment. What is more, the deformation
of the CN may also bring about an anisotropic angular distribution
pattern of the $\gamma$ ray, which implies that investigation of the
anisotropic angular distribution of GDR emission may also be a
promising way to obtain information on the deformation of the
composite system, which deserves further investigation.

\begin{figure}
\includegraphics[scale=0.19]{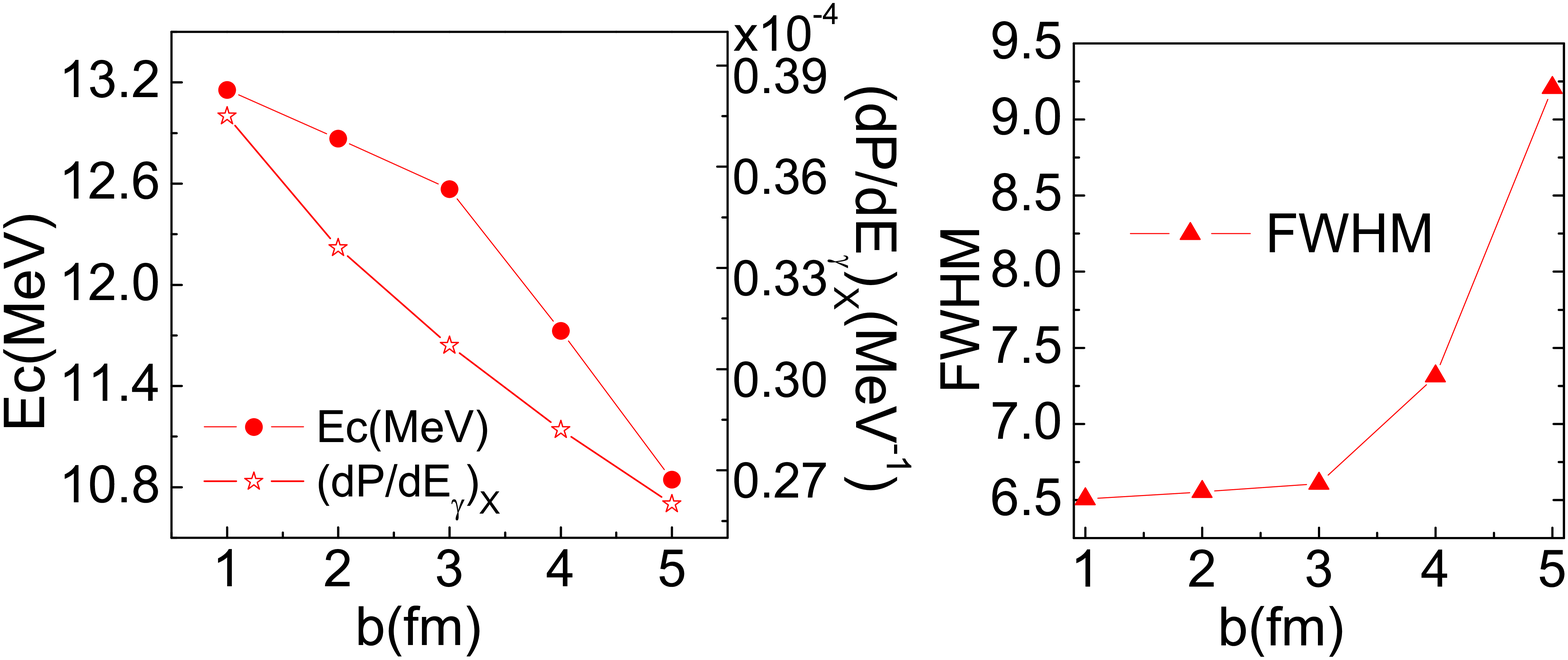}
\vspace{-0.1truein} \caption{\footnotesize (Color online) Left:
Central energy $E_{c}$ of the system $^{40}$Ca+$^{48}$Ca and
corresponding maximum $\gamma$-emission probability
$(dP/dE_{\gamma})_{x}$ as a function of $b$ with $E_{\rm beam} = 10$
MeV/nucleon and soft EOS parameters. Right: Corresponding variation
of the pre-equilibrium  GDR width of the system.}\label{Fig4}
\end{figure}

Dependences of the centroids of pre-equilibrium GDR $\gamma$-ray
spectra and maximum yields on the impact parameter are plotted in
the left panel in Fig. 4; both the central energy (i.e., peak
energy, denoted $E_{c}$) and the corresponding maximum $\gamma$-ray
yield, $(dP/dE_{\gamma})_{x}$, decrease monotonously with increasing
impact parameter. Meanwhile, the slower isospin equilibration
process for a large impact parameter may enhance the emission of
low-energy and multicomponent $\gamma$ rays, and this could explain
why the dynamical dipole emission width (denoted FWHM) increases
with the impact parameter, which is shown in the right panel in
Fig.\ref{Fig4}.

\begin{figure}
\includegraphics[scale=0.19]{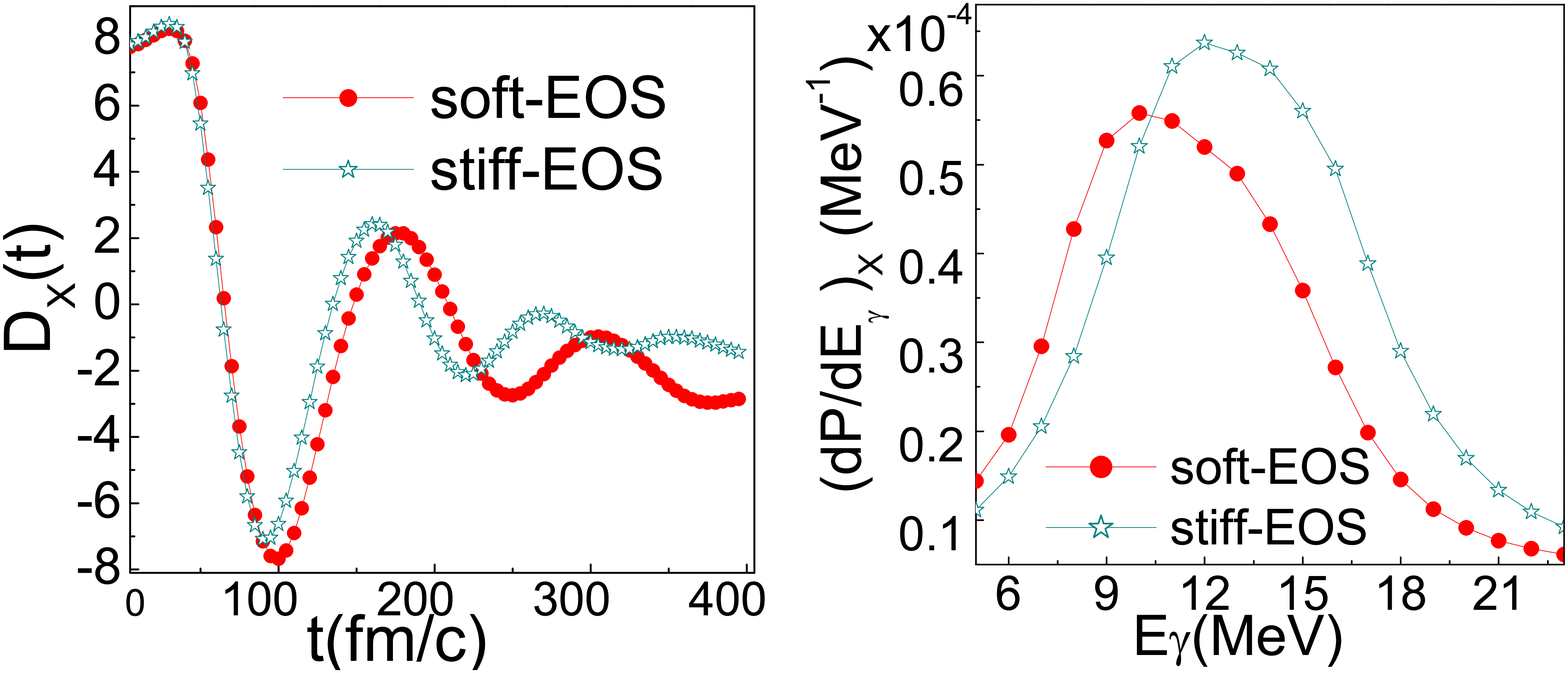}
\vspace{-0.1truein} \caption{\footnotesize (Color online) Left: The
giant dipole moment evolves with time for the $^{40}$Ca+$^{48}$Ca
system at $E_{\rm beam} = 10$ MeV/nucleon and $b=4$ fm;
Right: The corresponding $\gamma$-emission probability of the
system. Symbols are defined in the keys.}\label{Fig5}
\end{figure}

The sensitivity of the $\gamma$-ray yield to the stiffness of the
EOS is also studied by comparing the simulation with soft and stiff
Skyrme potentials.  The evolution of a giant dipole moment for the
$^{40}$Ca+$^{48}$Ca system, at $E_{\rm beam}$ = 10 MeV/nucleon and
$b$ = 4 fm, with different EOS parameters, soft EOS (filled circles)
and stiff EOS (open stars), are plotted in the left panel in
Fig.\ref{Fig5}; the corresponding $\gamma$-emission probability of
the system, in the right panel. As shown in the figure, the stiff
Skyrme potential generally leads to larger centroid and stronger
dynamical dipole $\gamma$ ray than the soft one.

\begin{figure}
\includegraphics[scale=0.19]{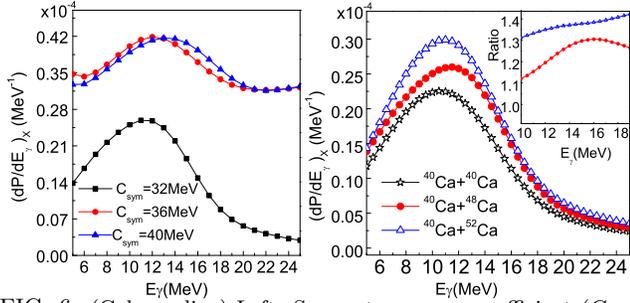}
\vspace{-0.2truein} \caption{\footnotesize (Color online) Left:
Symmetry energy coefficient ($C_{\rm sym}$) dependence of
$\gamma$-emission probability for the system $^{40}$Ca+$^{48}$Ca at
$E_{\rm beam} = 10$ MeV/nucleon and $b = 4$fm. Right:
$\gamma$-emission probability of systems $^{40}$Ca+$^{52}$Ca,
$^{40}$Ca+$^{48}$Ca and reference system $^{40}$Ca+$^{40}$Ca at
$E_{\rm beam} = 10$ MeV/nucleon and $b = 4$ fm with soft EOS
($C_{\rm sym}= 32$ MeV). Inset: Ratio of $\gamma$ yields of systems
with a $^{48}$Ca target or $^{52}$Ca target to those with a
$^{40}$Ca target. Symbols are defined in the keys.}\label{Fig6}
\end{figure}

Furthermore, the correlation between the dynamical dipole emission
and the symmetry term of the EOS is also investigated by changing
the symmetry energy coefficient ($C_{\rm sym}$). In the left panel
in Fig.\ref{Fig6}, the spectrum of dynamical emission of the same
system is reported, and the calculation is performed in the same
situation except for the $C_{\rm sym}$. We can clearly see an
increased yield of $\gamma$ rays for larger $C_{\rm sym}$ values.

As mentioned before, another interesting aspect of the GDR emission
property is its dependence on the $N/Z$ asymmetry of the reaction
systems, which in fact is a direct result of Eq.(6). This
relationship, indeed, is of most importance for its association with
the symmetry term of the EOS. Presently, taking the charge and mass
symmetric system $^{40}$Ca+$^{40}$Ca as a reference system, whose
pre-equilibrium effect is not expected because of small initial
dipole moment, we performed calculations of dipole $\gamma$ emission
in $N/Z$ asymmetry systems, namely, $^{40}$Ca+$^{48}$Ca and
$^{40}$Ca+$^{52}$Ca. In the right panel in Fig. 6, we show our
results for $\gamma$ emission in the early 400 fm/$c$ at $E_{\rm
beam} = 10$ MeV/nucleon, with $b = 4$ fm and soft EOS. In the inset,
the ratios of $\gamma$ yields are also shown, where the ratio
between the $^{40}$Ca+$^{48}$Ca and the $^{40}$Ca+$^{40}$Ca systems
is shown by filled circles and the ratio between the
$^{40}$Ca+$^{52}$Ca and the $^{40}$Ca + $^{40}$Ca systems is denoted
by open triangles. From the figure, we find that the $N/Z$ asymmetry
systems $^{40}$Ca+$^{52}$Ca $(N/Z = 1.3)$ and $^{40}$Ca+$^{48}$Ca
$(N/Z=1.2)$ show a clear enhanced yield of the order of about 35\%
and 20\%, with respect to the symmetry system $^{40}$Ca+$^{40}$Ca
$(N/Z=1.0)$, consistent with results in Ref.\cite{Pa03} and
references therein.

Experimental observations in Refs.\cite{Th93}and \cite{Fl96}
demonstrated that statistical $\gamma$ emission is insensitive to a
particular reaction entrance channel when we select the same
phase-space region of the initial CN. Therefore, the evidenced extra
$\gamma$ yield can be explained by the dynamics of the compound
system formed by a target and projectile with a large $N/Z$
asymmetry. In contrast, the equilibrium GDR width and the peak
energy are rather insensitive to isospin asymmetry \cite{Wa00}.

In summary, we have applied the QMD model to investigate
pre-equilibrium giant dipole oscillations (the dynamical dipole) and
made systematical calculations for the low-energy reactions of
$^{40}$Ca+Ca isotopes. The results show that dynamical dipole
$\gamma$ radiation is a good probe for gathering important
information on the early stage of fusion, in particular, the charge
asymmetry entrance channel. The dependences of the beam energy and
impact parameter of the dipole resonance are discussed. The extra
yield of pre-equilibrium GDR emission for systems with with a large
$N/Z$ asymmetry suggests a strong dependence of dynamical emission
on isospin asymmetry of the CN or, in other words, on the symmetry
term of the EOS of nuclear matter. Meanwhile, the sensitivities of
dynamical spectrum properties to the mean field potential and its
symmetry energy portion indicate that detailed study of dynamical
dipole $\gamma$ radiation can be used as a probe to study the EOS
around the normal nuclear density.

This work was supported in part by the National Natural Science
Foundation of China under Contract Nos.  10775167, 10979074,
10875167, and 10747163, the Major State Basic Research Development
Program in China under Contract No. 2007CB815004, the Chinese
Academy of Science Foundation under Grant No. CXJJ-216£¬ and the
Shanghai Development Foundation for Science and Technology under
Contract No. 09JC1416800.


\end{document}